\documentclass[fleqn,usenatbib]{mnras}

\usepackage{newtxtext,newtxmath}

\usepackage[T1]{fontenc}

\DeclareRobustCommand{\VAN}[3]{#2}
\let\VANthebibliography\thebibliography
\def\thebibliography{\DeclareRobustCommand{\VAN}[3]{##3}\VANthebibliography}

\usepackage{graphicx}	
\usepackage{amsmath}	

\usepackage{amssymb}	

\title[The radiative properties of LHAASO J1908+0621]{Investigating the radiative properties of LHAASO J1908+0621}

\author[Wu et al.]{
Keyao Wu,$^{1}$
Liancheng Zhou,$^{1}$
Yunlu Gong,$^{1}$
and Jun Fang$^{1,2}$\thanks{E-mail: fangjun@ynu.edu.cn }
\\
$^{1}$Key Laboratory of Astroparticle Physics of Yunnan Province, Yunnan University, Kunming 650091,
China;\\
$^{2}$Department of Astronomy, Yunnan University, Kunming 650091, China; fangjun@ynu.edu.cn\\
Received 20xx month day; accepted 20xx month day
}

\date{Accepted XXX. Received YYY; in original form ZZZ}

\pubyear{2015}

\begin{document}
\label{firstpage}
\pagerange{\pageref{firstpage}--\pageref{lastpage}}
\maketitle

\begin{abstract}
LHAASO J1908+0621 has been recently detected as a source emitting $\gamma$-rays with energies above 100 TeV, and the multiband observations show that a break around 1 TeV appears in the $\gamma$-ray spectrum. We have reanalyzed the GeV $\gamma$-ray properties for the 100 TeV source using 14 years of data recorded by the Fermi Large Area Telescope (Fermi-LAT). The spectrum in the energy range range 30-500 GeV has an index of 1.50 $\pm$ 0.26, which is much smaller than that detected in the TeV $\gamma$-rays. Additionally, the radiation properties of this source are investigated based on a one-zone time-dependent model. In the model,  LHAASO J1908+0621 is associated with a pulsar wind nebula (PWN) powered by the pulsar PSR J1907$+$0602. High-energy particles composed of electrons and positrons are injected into the nebula. Multiband nonthermal emission is produced via synchrotron radiation and inverse Compton scattering (ICS). Taking the effect of radiative energy losses and adiabatic cooling into account, the spectral energy distribution from the model with a broken power-law for the distribution of the injected particles can explain the detected fluxes in the $\gamma$-ray bands. The results support that LHAASO J1908+0621 originates from the PWN powered by PSR J1907$+$0602, and the $\gamma$-rays with energy above 100 TeV are produced by the electrons/positrons in the nebula via ICS.
\end{abstract}




\section{Introduction}
Multiple galactic very high-energy (VHE; $E > 100$ GeV) $\gamma$-ray  sources have been discovered \citep{2019arXiv190908609H,2018A&A...612A...1H,2021Natur.594...33C}, and PWNe are potential galactic sources emitting $\gamma$-rays with energies above 100 TeV \citep{2020MNRAS.498.4901F,2021RAA....21..286W,2022NewA...9001669Y}. A prominent source is the Crab nebula, from which photons with energies up to 1.4 PeV have been identified by the Large High Air Shower Observatory (LHAASO). Pulsars can blow high-speed magnetized relativistic pulsar winds, which are usually supersonic relative to its surrounding medium. Those relativistic winds can produce termination shocks in the outer boundary of the nebulae. At the shocks, charged particles can be accelerated to relativistic energies, and they can emit nonthermal emission from radio to $\gamma$-rays through synchrotron radiation and ICS. 

LHAASO is one of China's crucial national science and technology infrastructures. It has discovered more than a dozen VHE cosmic accelerators within the Milky Way galaxy \citep{2021Natur.594...33C,2021JPhG...48f0501A}. LHAASO J1908+0621 is one of the sources with radiation over 100 TeV detected by LHAASO. It is spatially coincident with MGRO J1908+06 \citep{2017ApJ...843...40A}, HESS J1908+063, eHAWC J1907+063 \citep{2021JCAP...08..010S} in TeV $\gamma$-rays, and  4FGL J1906.2+0631 in GeV $\gamma$-rays \citep{2020ApJS..247...33A}. MGRO J1908+06, discovered by the MILAGRO team during a 7-year survey of the northern hemisphere, is located in the Galactic plane \citep{2007ApJ...664L..91A}. The search for its multiband counterpart in radio and X-rays has not been very successful \citep{2007ATel.1251....1K,2010ApJ...711...64A,2015ICRC...34..743P,2020MNRAS.491.5732D}. But it was confirmed as an extended VHE source by many high-energy observation experiments, including the High Energy Stereoscopic System (HESS) \citep{2009A&A...499..723A}, the Very Energetic Radiation Imaging Telescope Array System  (VERITAS) \citep{2008AIPC.1085..301W,2014ApJ...787..166A}, the Astrophysical Radiation with Ground based Observatory at YangBaJing (ARGO-YBJ) \citep{2012ApJ...760..110B}, the High-Altitude Water Cherenkov (HAWC) \citep{2017ApJ...843...40A} and LHAASO \citep{2021Natur.594...33C}.  \citet{2021ApJ...913L..33L} firstly identified an extended GeV source as the counterpart of MGROJ1908+06 and discovered potentially relevant molecular clouds (MCs). It is widely indicated that the source has a hard spectral index, while ARGO reports a slightly softer spectral index of $2.54 \pm 0.36$ \citep{2012ApJ...760..110B}.

Several objects could be the counterparts of LHAASO J1908+0621, including a middle-aged radio supernova remnant SNR G40.5-0.5 and several pulsars. The nearby shell-type SNR G40.5-0.5 is estimated to have an age of $20 - 40$ kyrs \citep{1980A&A....92...47D,2009BASI...37...45G}. The distance of this SNR G40.5-0.5 is uncertain. Using the $\Sigma$-D relation, where $\Sigma$ is the surface brightness and D is the diameter of the SNR, \citep{1980A&A....92...47D} estimated a distance of $5.5 - 8.5$ kpc, and \citep{2020MNRAS.491.5732D} considered ~8.7 kpc as the most likely distance to G40.5-0.5 by CO observations. There are also three pulsars near LHAASO J1908+0621. The most powerful one is the radio faint pulsar, PSR J1907+0602, found by the Fermi-LAT in the GeV energy range \citep{2010ApJ...711...64A}. It is located south of MGRO J1908+06, slightly off the peak of $\gamma$-ray excess counts. According to the ATNF pulsar database, this pulsar has a spin-down power ($\Dot{E}$) of $2.8\times10^{36} \,\mathrm{erg} \,\mathrm{s}^{-1}$, a characteristic age of 19.5 kyr, and an estimated distance  of 2.37 kpc \citep{2005AJ....129.1993M}. There are also two weak pulsars, PSR J1907+0631 and PSR J1905+0600. The spin-down power $\Dot{E}$ of PSR J1907+0631 is $5.3\times10^{35} \,\mathrm{erg} \,\mathrm{s}^{-1}$. It is located very close to the center of the SNR G40.5-0.5, and the estimated age is consistent with the one estimated for the remnant \citep{2017ApJ...834..137L}. \citet{2020MNRAS.491.5732D} show that, in principle, PSR J1907+0631 can power the entire TeV source, which needs a conversion efficiency of about $3\%$ for the rotational energy \citep{2007Ap&SS.309..197G}. The pulsar PSR J1905+0600 has a distance of 18 kpc and a spin-down power of $5.1\times10^{32} \,\mathrm{erg} \,\mathrm{s}^{-1}$, which is too weak to contribute to TeV radiation.

In this paper, we have analyzed the data observed by Fermi-LAT during the period 2008 August 4 to 2022 April 5. Based on the obtained Fermi-LAT data results in combination with the TeV $\gamma$-ray fluxes with HAWC and LHAASO, we use a one-zone time-dependent model for the multiband nonthermal emission of a PWN to investigate whether the detected $\gamma$-rays from LHAASO J1908+0621 can be generated by the particles powered by the pulsar, and the averaged magnetic field strength in the emission region can be constrained. The details of the model are illustrated in section \ref{model}, the Fermi-LAT data processing is indicated in section \ref{Fermi-LAT Data Analysis}, and the model results are shown in section \ref{Conclusions}. Finally, the summary and some discussions are indicated in section \ref{summary and discussion}.

\section{The model}\label{model}

\begin{figure*}
\centering
\begin{minipage}[t]{0.48\textwidth}
\centering
\includegraphics[height=8cm,width=8cm]{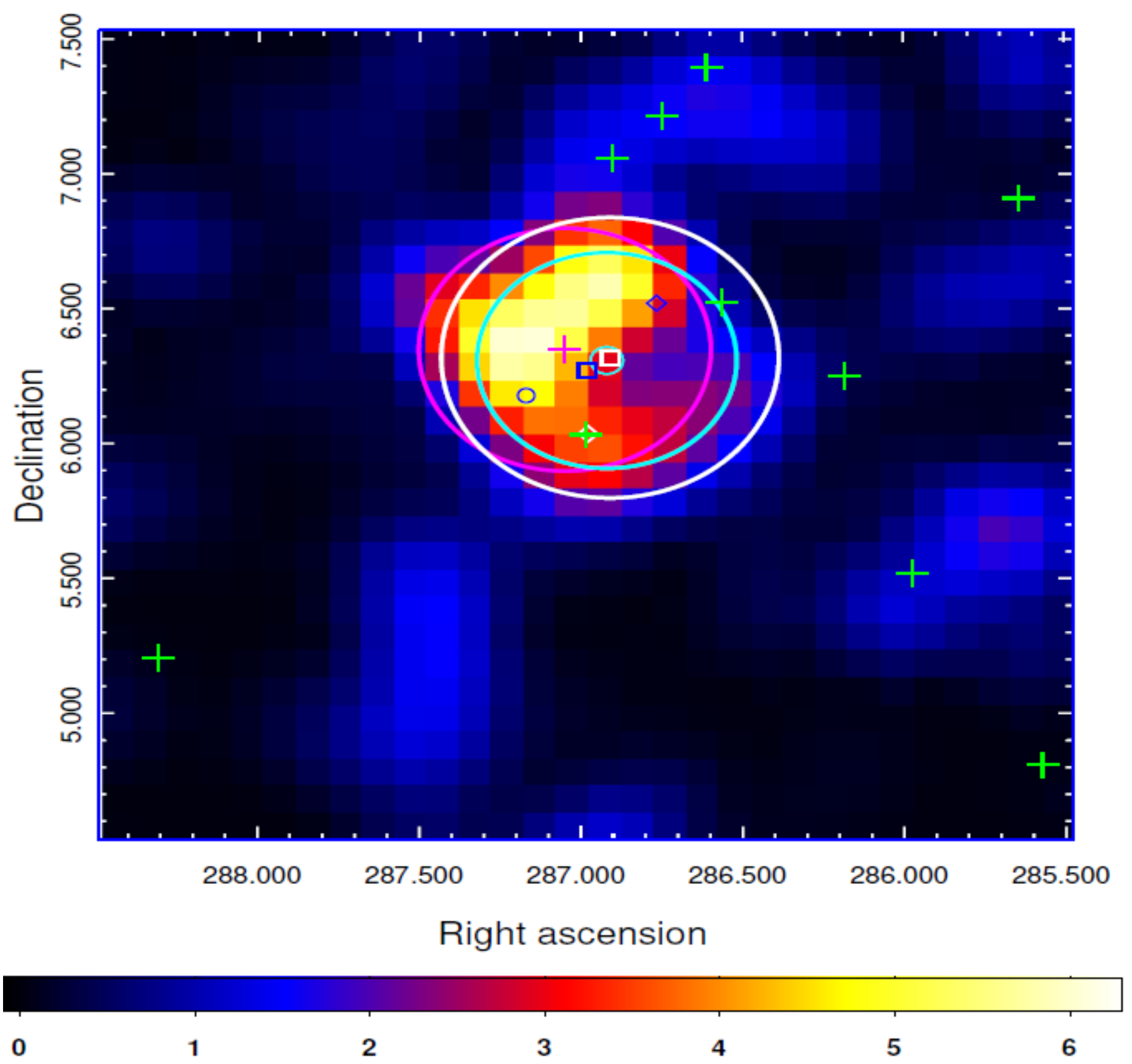}
\end{minipage}
\begin{minipage}[t]{0.48\textwidth}
\centering
\includegraphics[height=8cm,width=8cm]{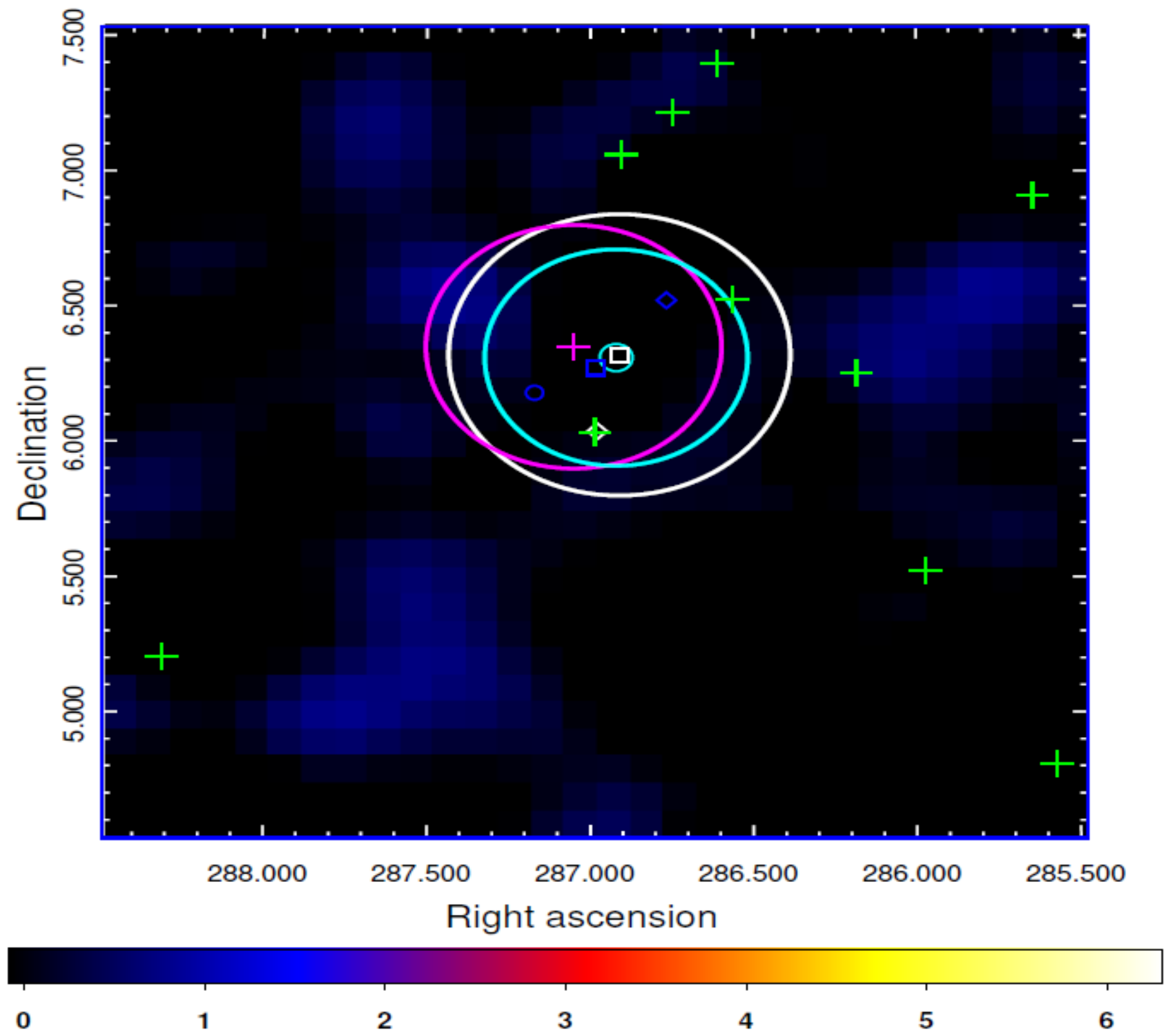}
\end{minipage}
\caption{Fermi-LAT Test Statistic (TS) map of SrcX region in the 30–500 GeV range. 
Left panel: TS map including all sources (green cross) from 4FGL catalog. Right panel: TS map after subtracting all sources containing SrcX. The large and small cyan circles show the Fermi optimal extension radius and the 68\% error of the best-fit position of SrcX, respectively. Magenta cross and circle indicate the LHAASO position (19.21$^\prime$ away from pulsar J1907+0602) and the 68\% contamination radius of disk morphology determined above 25 TeV, respectively. The white box and circle represent the position given by HAWC and its extension radius, respectively. PSR J1907+0602 is indicated with a white diamond and the possibly associated pulsar PSR J1907+0631 is shown with a blue diamond. MGRO J1908+06 is shown with a blue circle. The TS maps are smoothed with a Gaussian kernel of $0.6^{\circ}$.}
\label{Fig1}
\end{figure*}

We assume the emission region of the $\gamma$-ray source is a sphere containing high-energy electrons/positrons accelerated by the termination shock of a PWN, and the energy spectrum of particles at time $t$ can be derived from the equation \citep{2012MNRAS.427..415M,2010A&A...515A..20F,2014JHEAp...1...31T},
\begin{equation}
   \frac{\partial N(\gamma,t)}{\partial t}=-\frac{\partial}{\partial \gamma}[\Dot{\gamma}(\gamma,t)N(\gamma,t)]-\frac{N(\gamma,t)}{\tau(\gamma,t)}+Q(\gamma,t) ,
\end{equation} 
where the left side is the distribution of the particles over time, and the first term on the right represents the continuous energy change of particles due to energy losses. $\Dot{\gamma}(\gamma,t)$ is the sum of energy loss rate caused by synchrotron, ICS and adiabatic cooling \citep{2012MNRAS.427..415M}. $Q(\gamma,t)$ represents the number of particles injected per unit energy and per unit volume at $t$. $\tau(\gamma,t)$ is the escape time of the high-energy leptons leaving the nebula via Bohm diffusion \citep{2008ApJ...676.1210Z}. High-energy leptons can be continually injected into the PWN in various forms with a distribution of power law\citep{2009ApJ...703.2051G,1984ApJ...283..710K,1984ApJ...278..630R}
\begin{equation}
   Q(\gamma,t)=Q_0(t)\gamma^{\mathrm{-\alpha}} ,
\end{equation}
or broken power law \citep{2012MNRAS.427..415M,2014JHEAp...1...31T,2004A&A...422..609B}
\begin{equation}
   Q(\gamma,t)= Q_0(t)
   \begin{cases}
   (\frac{\gamma}{\gamma_\mathrm{b}})^{\mathrm{-\alpha_1}} \quad if \gamma \le \gamma_\mathrm{b} \\\\
   (\frac{\gamma}{\gamma_\mathrm{b}})^{\mathrm{-\alpha_\mathrm{2}}} \quad if \gamma_\mathrm{b} < \gamma \le \gamma_\mathrm{max}
   \end{cases} ,
\end{equation}
where $\gamma_\mathrm{b}$ is the break energy, the parameters $\alpha$, $\alpha_1$ and $\alpha_2$ are the spectral indices.
In order to ensure the confinement of the particles, the Larmor radius of the particles must be smaller than the termination shock radius, which leads to \citep{2009ASSL..357..451D}
\begin{equation}
   \gamma_{\mathrm{max}}=\frac{\epsilon e \kappa}{m_{\mathrm{e }}c^2}\sqrt{\eta\frac{L(t)}{c}} ,
\end{equation}
where $e$ is the electron charge,  $\epsilon$ is the fractional size of the shock radius, $\kappa=3$ is the magnetic compression ratio.  The normalized constant $Q_0(t)$ can be determined by
\begin{equation}
   (1-\eta)L(t)=\int \gamma m_e c^2 Q(\gamma,t) d\gamma ,
\end{equation}
where $\eta$ is the particle energy fraction. For a pulsar with a period of $P$ and a period derivative of $\Dot{P}$, the spin-down luminosity of the pulsar is \citep{1969ApJ...157..869G,2006ARA&A..44...17G}
\begin{equation}
   L(t)=4\pi^2 I\frac{\Dot{P}}{P^3}=L_0 \left( 1+\frac{t}{\tau_0} \right )^{-\frac{n+1}{n-1}} ,
\end{equation}
where $L_0$ is initial luminosity, $I=10^{45} \mathrm{g} \,\mathrm{cm}^2$ is moment of inertia of the pulsar, and $n$ is the breaking index. The initial spin-down time-scale of the pulsar with an age of $t_{\mathrm{age}}$ is \citep{2006ARA&A..44...17G}
\begin{equation}
   \tau_0=\frac{P_0}{(n-1)\Dot{P_0}}=\frac{2\tau_c}{n-1}-t_{\mathrm{age}} .
\end{equation}
$P_0$ and $\Dot{P_{\mathrm{0}}}$ is the initial period of the pulsar and its first derivative, respectively, and $\tau_{\mathrm{c}} = P/2\Dot{{P}}$ is the characteristic age of the pulsar.

In the PWN, the magnetic field strength $B(t)$ results from solving the equation, \citep{2010ApJ...715.1248T,2012MNRAS.427..415M,2014JHEAp...1...31T}
\begin{equation}
   \int_{0}^{t} \eta L(t^{'})\mathrm{d}t^{'}=\frac{4\pi}{3}R_{\mathrm{PWN}}^{3}(t)\frac{B^2(t)}{8\pi} ,
\end{equation}
therefore, using the above $L(t)$ correlation equation and solving for the field, we obtain
the magnetic field strength in the PWN is
\begin{equation}
    B(t) = \sqrt{\frac{3(n-1) \eta L_{\mathbf{0}} \tau_{\mathrm{0}}} {R_{\mathrm{PWN}}^{3}(t)} \left[ 1-\left(1+\frac{t}{\tau_{\mathrm{0}}}\right)^{-\frac{2}{n-1}}\right] } ,
\end{equation}
Assuming that the PWN is in the free expansion phase, its radius is \citep{2001A&A...380..309V,2003A&A...404..939V,2014JHEAp...1...31T}
\begin{equation}
    R_{\mathrm{PWN}}(t) = C(\frac{L_0t}{E_0})^{\frac{1}{5}}\sqrt{(\frac{10E_0}{3 M_{\mathrm{ej}}})} t,
\end{equation}
where 
\begin{equation}
    C = \left(\frac{6}{15(\gamma_{\mathrm{PWN}}-1)}+\frac{289}{240}\right)^{-\frac{1}{5}} ,
\end{equation}
with $\gamma_{\mathrm{PWN}} = 4/3$, and $E_0$ is kinetic energy of the supernova ejecta, $M_{\mathrm{ej}}$ is ejected mass. The parameters $\epsilon$, $E_0$, and $M_{\mathrm{ej}}$ are random assumed to be  the typical values $0.4$, $10^{51} \,\mathrm{erg}$ and $9.5 M_{\odot}$ for the PWN in this paper.

\section{FERMI-LAT Data Analysis}\label{Fermi-LAT Data Analysis}
\subsection{Data Reduction}

We use the available software \texttt{Fermipy} \citep[][]{2017ICRC...35..824W} with the binned likelihood analysis method to perform the analysis of GeV emission towards LHAASO J1908+0621. The latest Pass 8 version of the data with “Source” class (evclass=128) “FRONT+BACK” type (evtype=3) events and the instrumental response function \texttt{P8R3$\_$SOURCE$\_$V3} are adopted, recorded from 2008 August 4 to 2022 April 5 ({\it Fermi} mission elapsed time 239557417s - 670841240s). We also use the filter expression \texttt{(DATA\_QUAL>0)$\&\&$(LAT\_CONFIG}==1) to select good quality time data. To restrict the contamination from $\gamma$-ray from Earth’s limb, the maximum zenith angle is set to $90^\circ$. The region of interest (ROI) is selected as a $10^\circ \times 10^\circ$ square region centered at the position of 4FGL J1907.9+0602. Since the $\gamma$-ray emission region overlaps with the bright pulsar PSR J1907+0602, we only consider the energy range of 30 to 500 GeV to prevent the contamination from the pulsar.

In our background model, we included all sources within an ROI of $20^\circ$ around LHAASO J1908+0621 from the Fermi Large Area Telescope (Fermi-LAT) Fourth Source Catalog \citep[4FGL; ][]{2020ApJS..247...33A}. All the normalization factors and the spectral indices from sources within $5^\circ$ were set as free parameters. The normalization parameters of the two background templates from the Galactic diffuse emission (\texttt{gll$\_$iem$\_$v07.fits}) and the isotropic extragalactic emission (\texttt{iso$\_$P8R3$\_$SOURCE$\_$V3$\_$v1.txt}) were also set as free.

\subsection{Spectral Data Analysis}

\begin{figure}
\includegraphics[width=\columnwidth]{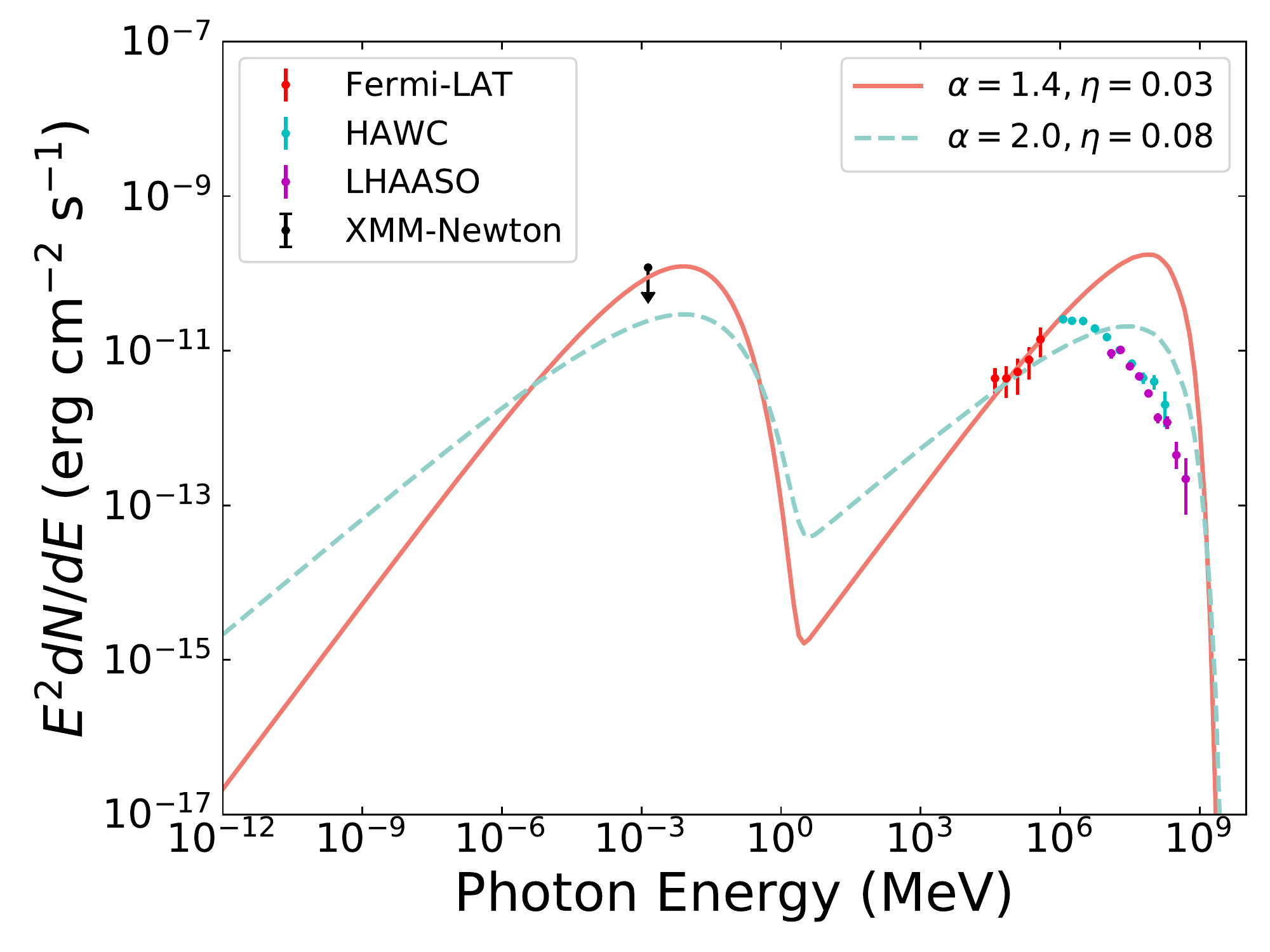}
    \caption{Comparison of the resulting SED from model with the detected fluxes associated with LHAASO J1908+0621. The observed fluxes of XMM-Newton \citep{2021ApJ...913L..33L}, Fermi-LAT, HAWC \citep{2020PhRvL.124b1102A} and LHAASO \citep{2021Natur.594...33C} are shown in the figure. }
    \label{fig:power-law}
\end{figure}

We first created a $3.0^\circ \times 3.0^\circ$ test statistic (TS) map centered at the position of 4FGL J1907.9+0602 in the 30–500 GeV energy band. From the left panel of Fig.~\ref{Fig1}, it can be found that there is significant $\gamma$-ray emission in the direction of LHAASO J1908+0621. This is consistent with the results found by \citet{2021ApJ...913L..33L} in the 30 to 500 GeV band. However, we also consider the case after subtracting the point source, and the results show that there is no significant residual emission at the target location, as shown in the right panel of Fig.~\ref{Fig1}. The above analysis suggests that $\gamma$-ray emission is likely to come from the target area. Then, the best-fit position (R.A., decl. = 286.92$^\circ$, 6.31$^\circ$, with 1$\sigma$ error radius of 0.05$^\circ$) from the $\gamma$-ray source calculated with \texttt{gtfindsrc} according to the position (R.A. = 286.88$^\circ$, decl. = 6.29$^\circ$) given by \citet{2021ApJ...913L..33L}. Hereafter, the best-fit position (marked as SrcX) of the $\gamma$-ray source is used for the following analysis.

We check whether SrcX is an extended source using the uniform disk models with different radii values ($0.3^\circ$, $0.4^\circ$, $0.5^\circ$, and $0.6^\circ$). The TS values for the different spatial templates are 32.69, 47.58, 42.52, and 33.69, respectively. We also adopted the Akaike information criterion \citep[AIC;][]{1974ITAC...19..716A} to chose the model that better describes SrcX from point source and uniform disk. It can be expressed as AIC = 2k $-$ 2 $\ln\mathcal{L}$, where k is the number of the free parameters in the model (for a point source k = 4 and for a uniform disk k = 5) and $\mathcal{L}$ is the maximum likelihood obtained in the fit. Then we got the significantly improved $\Delta$ AIC = ${AIC}_\mathrm{point}$ - ${AIC}_\mathrm{ext}$ $\approx$ 45.58, which the ${AIC}_\mathrm{point}$ and ${AIC}_\mathrm{ext}$ represent the AIC values of the point source and extended source respectively. Thus, we used the uniform disk with a radius of $0.4^\circ$ to describe the $\gamma$-ray emission of SrcX in the spectral analysis. Furthermore, the spectral index and flux of SrcX in the energy band of 30–500 GeV are fitted to be $1.50 \pm 0.26 \,\mathrm{and} (2.02 \pm 0.44) \times 10^{-11} \,\mathrm{erg} \,\mathrm{cm}^{-2} \,\mathrm{s}^{-1} $, respectively. We divided the data into five equal logarithmic energy bins between 30 and 500 GeV to calculate the SED of SrcX. The binned likelihood analysis method is reused to fit each energy bin. The SED of SrcX is shown by the red dot in Fig.~\ref{fig:J-1907}.

\section{Results}\label{Conclusions}

\begin{table*}
    \centering
	\caption{The different parameters for the models.}
	\label{tab:table1}
	\resizebox{0.8\textwidth}{!}{\begin{tabular}{lccccccccc|c}	 
    	\hline
		model & A & B & C & D & E & F & G & H & I & J\\
		\hline
		$\alpha_1$ & 1.4 & 2.0 & 1.0 & 1.0 & 1.0 & 1.0 & 1.3 & 1.5 & 1.0 & 1.0 \\
    	$\alpha_2$ & - & - & 2.9 & 2.9 & 2.9 & 2.9 & 2.9 & 2.9 & 2.5 & 2.0  \\
     	$\eta$ & 0.03 & 0.08 & 0.42 & 0.05 & 0.25 & 0.50 & 0.42 & 0.42 & 0.42 & 0.42\\
     	$\chi^2$ & 21832 & 14296 & 49 & 19710 & 1390 & 84 & 60 & 172 & 1328  & 10407  \\
		
		\hline
	\end{tabular}}
\end{table*}

\begin{figure}
\includegraphics[width=\columnwidth]{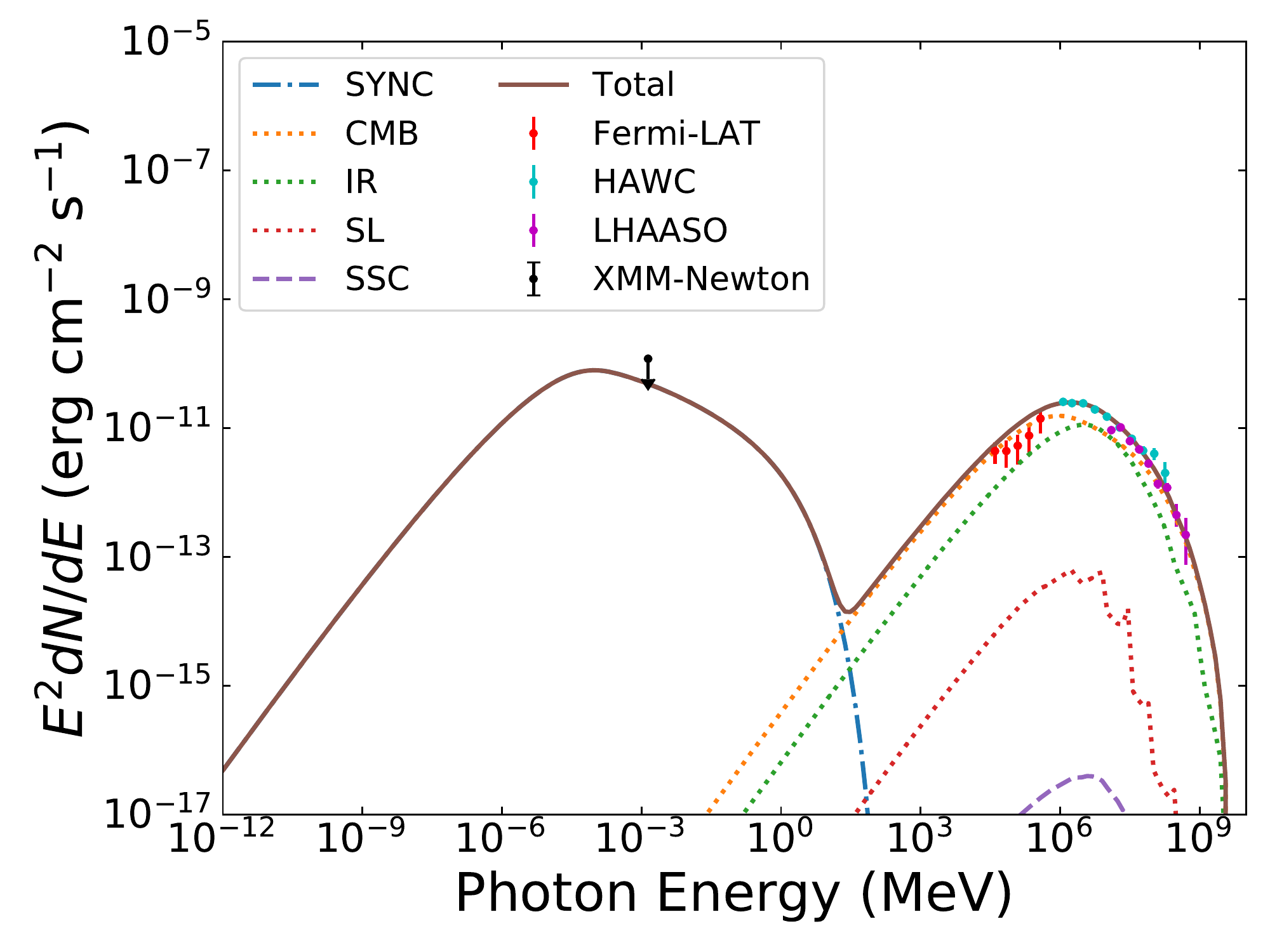}
    \caption{Comparison of the resulting SED from model C with the detected fluxes associated with LHAASO J1908+0621. The total emission (solid line) consists of sychrotron (dashdotted line) radiation and ICS (dotted lines). The observed fluxes of XMM-Newton \citep{2021ApJ...913L..33L}, Fermi-LAT, HAWC \citep{2020PhRvL.124b1102A} and LHAASO \citep{2021Natur.594...33C} are shown in the figure. }
    \label{fig:J-1907}
\end{figure}

\begin{figure}
\includegraphics[width=\columnwidth]{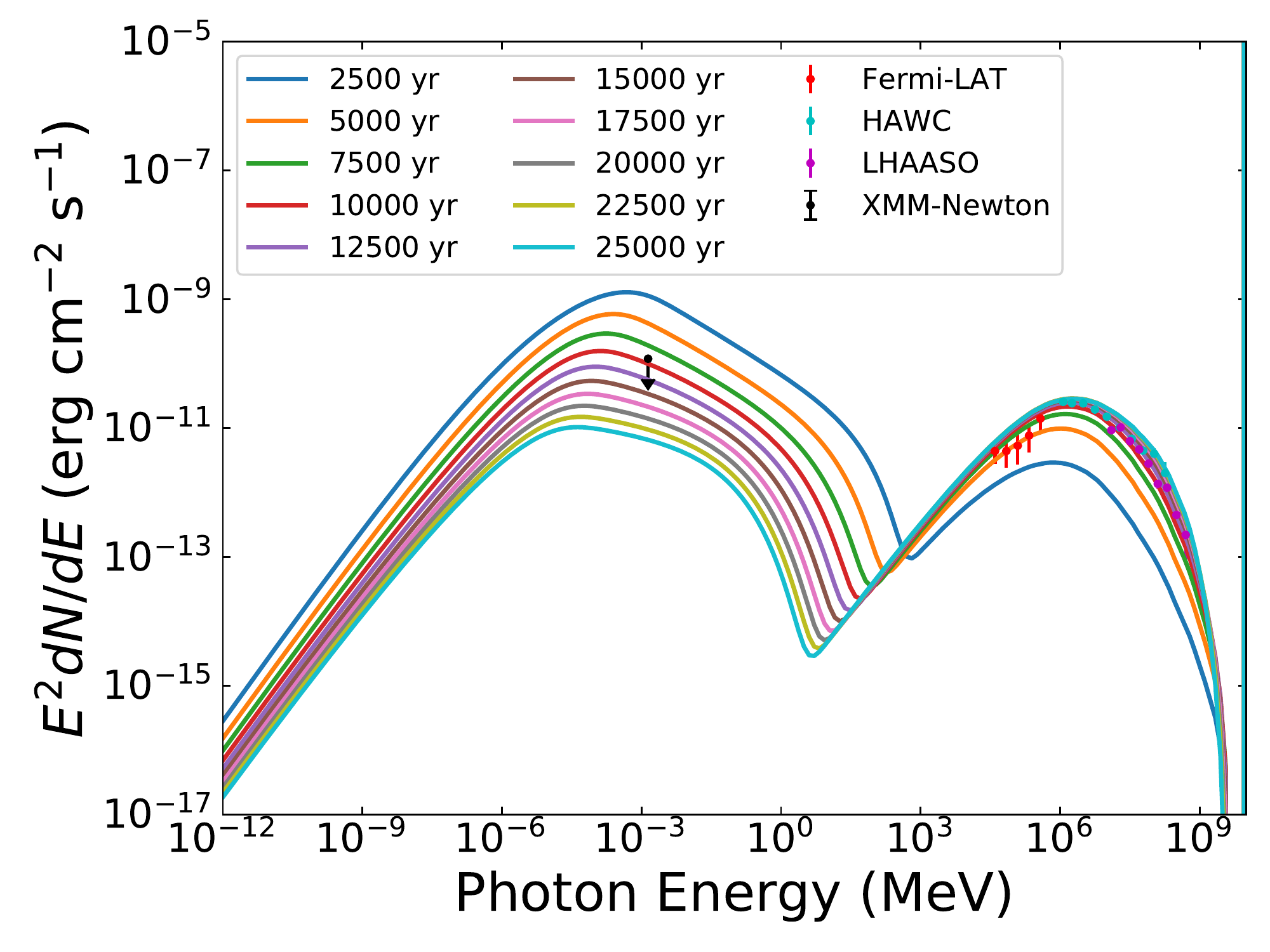}
    \caption{Multi-band SEDs of LHAASO J1908+0621 at different times from model C. The references for the detected fluxes are the same as Fig.~\ref{fig:J-1907} }
    \label{fig:J1907dw}
\end{figure}

\begin{figure}
\includegraphics[width=\columnwidth]{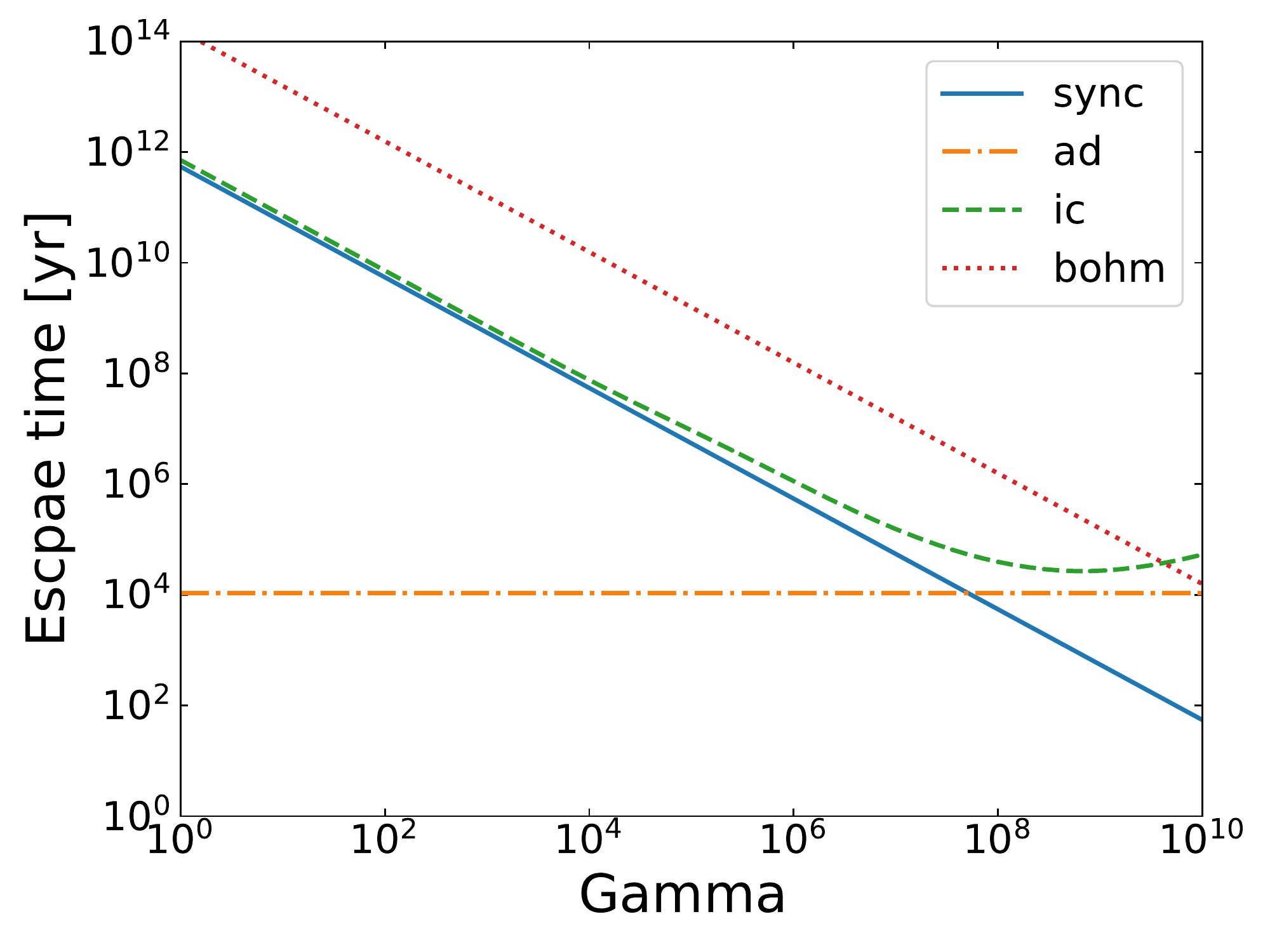}
    \caption{The time scales for the energy loss of synchrotron radiation (solid line), adiabatic loss (dashdotted line), ICS (dashed line) and the escape due to Bohm diffusion (dotted line) in the best-fit model at $t_{\mathrm{age}} = 13000$yr, respectively.}
    \label{fig:1907escapetime}
\end{figure}

\begin{figure*}
    \centering
 
    \begin{minipage}{0.495\linewidth}
     \centering
     \includegraphics[width=1.0\linewidth]{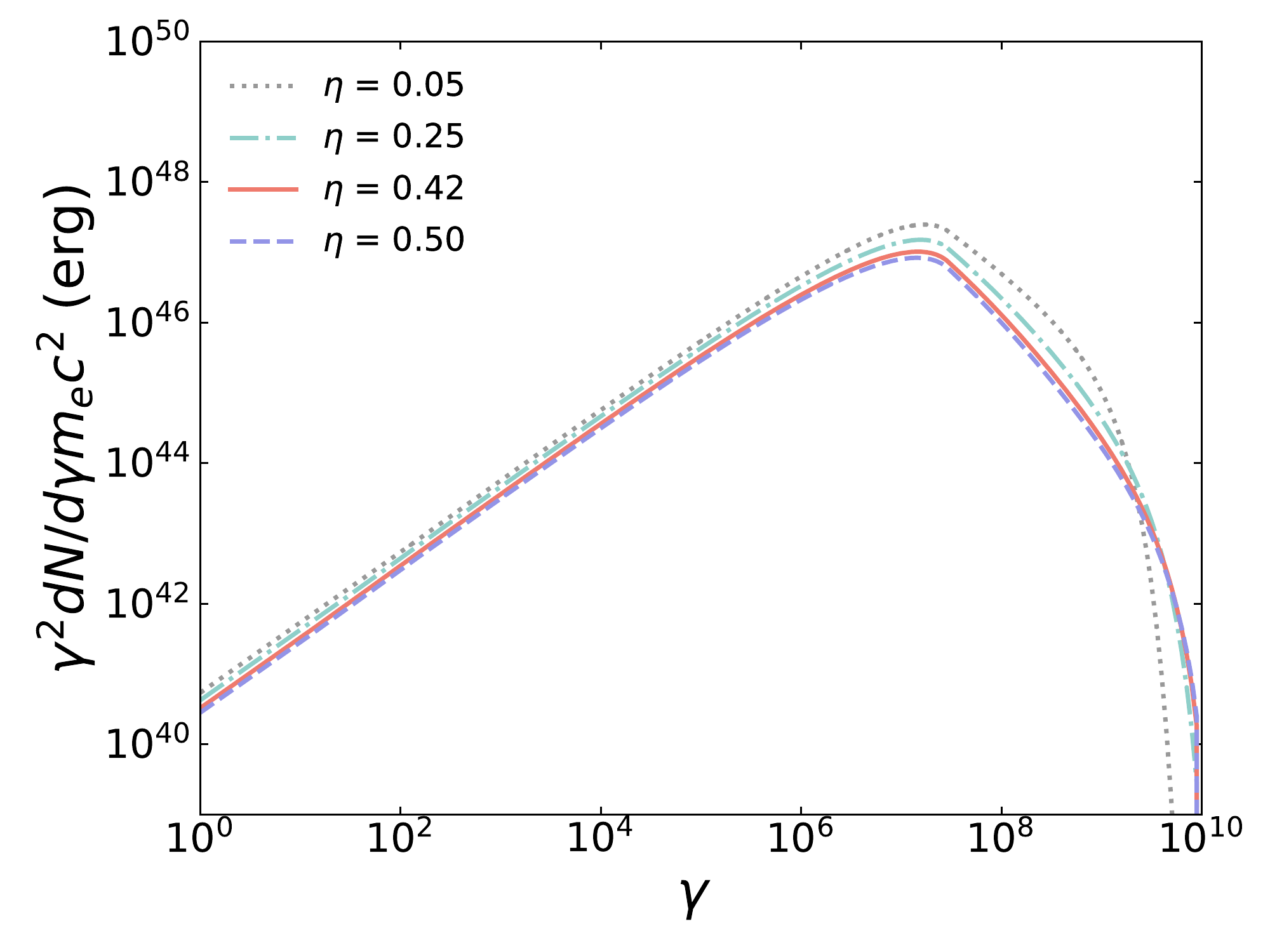}
    \end{minipage}    
    \begin{minipage}{0.495\linewidth}
     \centering
     \includegraphics[width=1.0\linewidth]{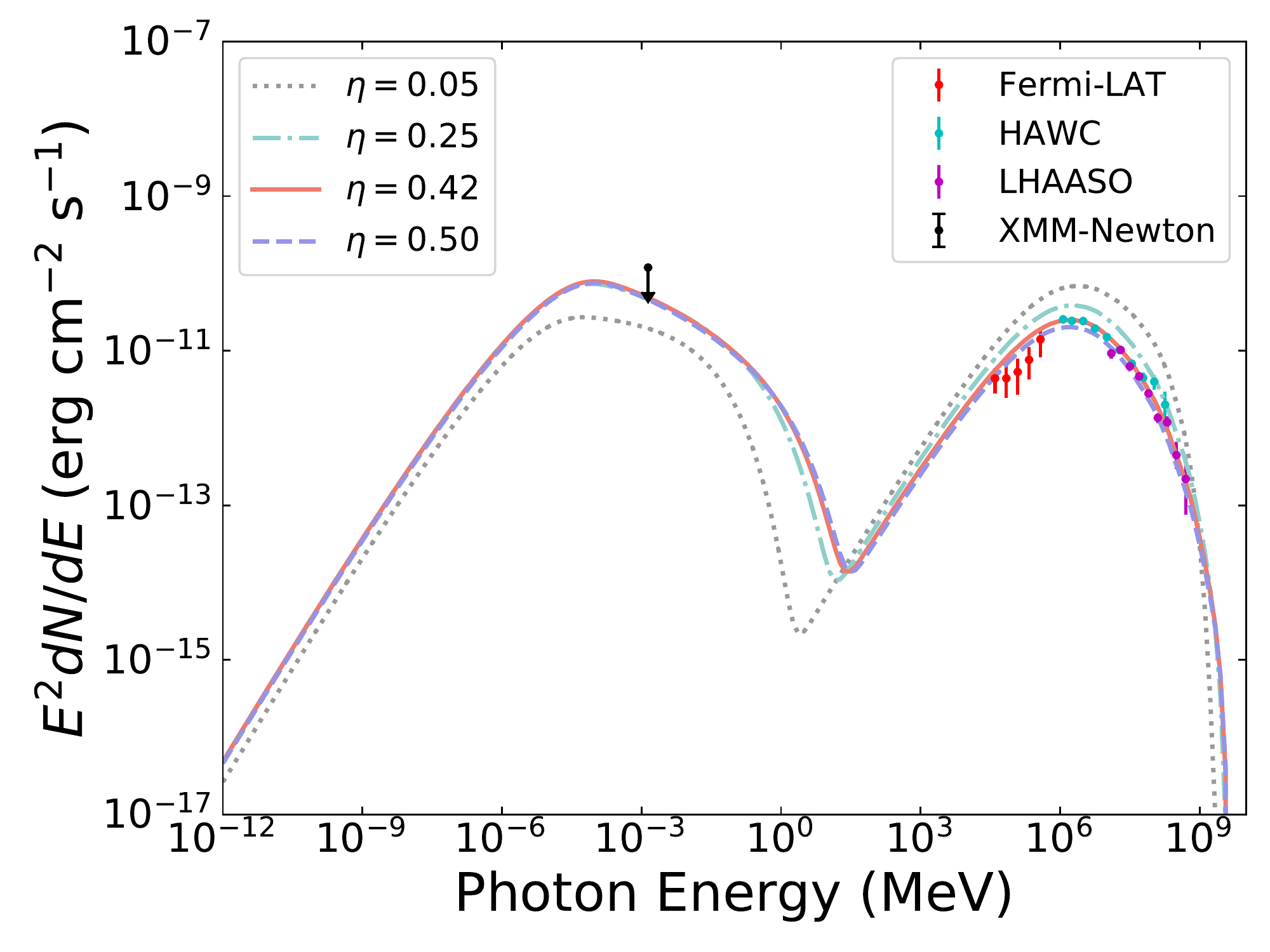}
    \end{minipage}

    \caption{ The particle distribution of electrons/positrons in the nebula for models C (the solid line), D (the dotted line), E (the dash dotted line) and F (the dashed line) (left panel) and the corresponding SEDs (right panel). The references on the detected fluxes are the same as Fig.~\ref{fig:J-1907}.}
    \label{fig:qgamma-eta}
\end{figure*}

\begin{figure*}
    \centering
 
    \begin{minipage}{0.495\linewidth}
     \centering
     \includegraphics[width=1.0\linewidth]{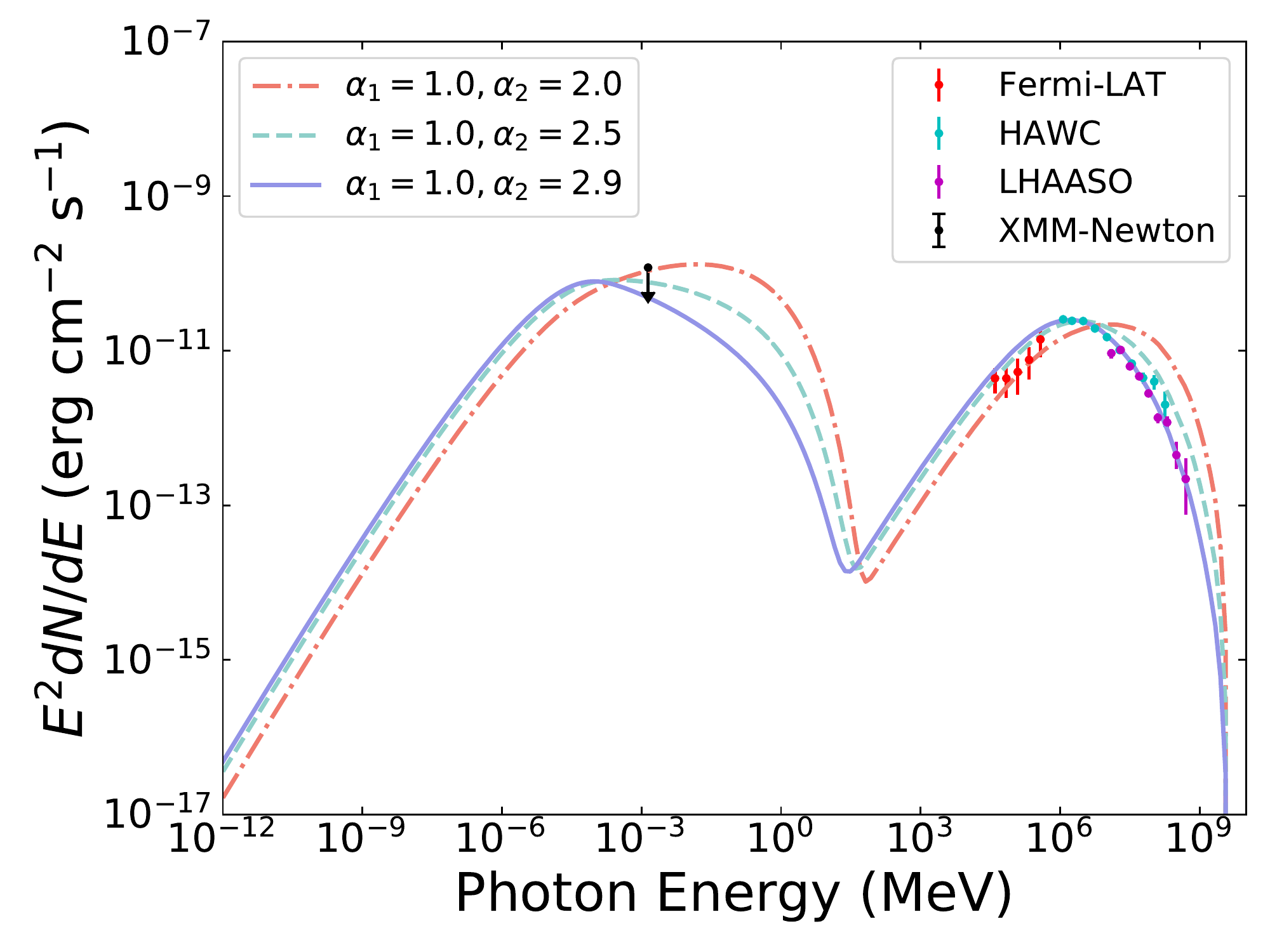}
    \end{minipage}    
    \begin{minipage}{0.495\linewidth}
     \centering
     \includegraphics[width=1.0\linewidth]{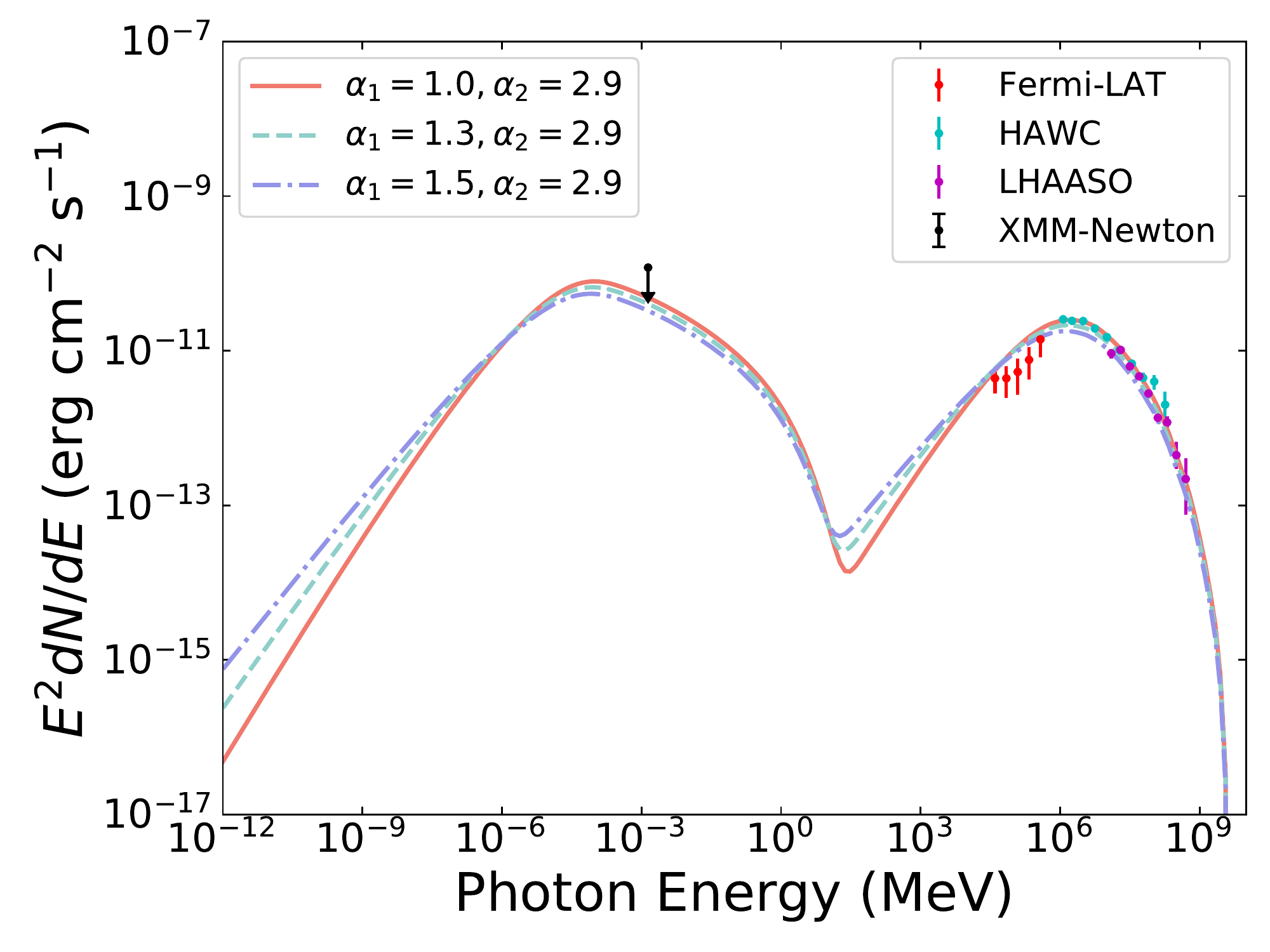}
    \end{minipage}

    \caption{The resulting SEDs from the different models. The references on the detected fluxes are the same as Fig.~\ref{fig:J-1907}}
    \label{fig:alpha}
\end{figure*}
In the one-zone time-dependent model, the multiband nonthermal emission consists of two components from radio to TeV $\gamma$-rays. The lower-energy emission is produced by synchrotron radiation from the electrons/positrons accelerated in the PWN. The GeV and TeV $\gamma$-rays are produced by ICS off the soft photons, including the cosmic microwave background (CMB), the infrared (IR) radiation, and star light (SL). The temperatures and the energy densities for the three components are $T_{\mathrm{IR}} = 2.7$ K, $U_{\mathrm{IR}} = 0.25 \,\mathrm{eV} \,\mathrm{cm}^{-3}$, $T_{\mathrm{IR}} = 20$ K, $U_{\mathrm{IR}} = 0.30 \,\mathrm{eV} \,\mathrm{cm}^{-3}$, $T_{\mathrm{SL}} = 5000$ K, $U_{\mathrm{SL}} = 0.30 \,\mathrm{eV} \,\mathrm{cm}^{-3}$ \citep{2017ApJ...846...67P}.  In addition, a self-synchrotron Compton (SSC) is also taken into account, but it is generally insignificant for the PWN.

The pulsar PSR J1907 + 0602 has a period  $P = 0.107$ s, a period derivative of $\Dot{P} = 8.68 \times 10^{-14} \mathrm{s}\, \mathrm{s}^{-1}$, which lead to a characteristic age $\tau_c = 1.95 \times 10^4$ yr and a spin-down luminosity of $L = 2.80 \times 10^{36} \,\mathrm{erg} \,\mathrm{s}^{-1}$ \citep{2021Natur.594...33C,2021JCAP...08..010S}. We also adopt a distance to the pulsar of 2.4 kpc and a braking index of $n=3$ when performing the fit.  Since the age of the pulsar is unknown, we first explore the possibility of different ages for the nebula, and an age of $T_{\mathrm{age}} = 13000$ yr is used to produce a flux consistent with the detected one in the $\gamma$-ray band.   For this age, the initial spin-down time-scale is $\tau_0 = 6.51 \times 10^3$ yr, and the initial luminosity is $L_0 = 2.51\times10^{37} \,\mathrm{erg} \,\mathrm{s}^{-1}$.  The indexes in the distribution of the injected particles, $\alpha_1$, $\alpha_2$, the energy break $\gamma_b$, and  $\eta$ are set as free parameters to reproduce a resulting SED consistent with the multiband observations.  

We first assume that the particles with a power law spectrum of energy distribution are continuously injected into the nebula and investigate whether the resulting SED of the nebula can be reconciled with the detected fluxes. Fig.~\ref{fig:power-law} shows the resulting SEDs with $\alpha=1.4$ (model A) and $2.0$ (model B). With a power-law distribution for the injected particles, the detected fluxes above 1 TeV detected by HAWC and LHAASO can not be well reproduced with the model.

Broken power law spectra for the particles injected into PWNe are widely used to reproduce SEDs consistent with multi-band detected fluxes. Assuming the electrons/positrons injected into the PWN of  PSR J1907+0602 have a broken power law spectrum, as indicated in Table 1, different models with various $\alpha_1$, $\alpha_2$, $\gamma_b$ and $\eta$ are chosen to seek appropriate parameters to reproduce a SED consistent with  the fluxes obtained by XMM-Newton, Fermi-LAT, HAWC and LHAASO for the nebula. As shown in Fig.~\ref{fig:J-1907}, in model C with $\alpha_1 = 1.0$, $\alpha_2 = 2.9$, $\gamma_b=2.8 \times 10^7$, and $\eta = 0.42$,  the detected $\gamma$-ray fluxes are well reproduced, and the $\gamma$-rays from associated with LHAASO J1908$+$0621 can be explained as the photons produced by the electrons/positrons via ICS.  Fig.~\ref{fig:J1907dw} indicates the evolution of the SED from model C over time. The magnetic field strength in the nebula decreases with time due to the expansion of the PWN, which leads to the attenuation of the synchrotron radiation. After $t\sim10^4\,\mathrm{yr}$, the radiation from the ICS is stabilized due to the balance between the injection and the energy losses of the particles.

In Fig.~\ref{fig:1907escapetime}, the cooling time scales of the electrons/positrons at $T_{\mathrm{age}}$ due to the synchrotron radiation, the adiabatic loss, the ICS and the escape for the  Bohm diffusion over $\gamma$ in model C are indicated. At $t=T_{\mathrm{age}} = 13000$ yr, the magnetic field strength of the nebula is $B=6.7 \,\mu$G. For $\gamma<10^8$, the main cooling process of the particles is adiabatic loss. For particles with $\gamma>10^8$, the energy loss due to synchrotron radiation is the most prominent one compared with ICS and Bohm diffusion.

Fig.~\ref{fig:qgamma-eta} shows the particle spectra and the resulting SEDs for different $\eta$ with $\alpha_1 = 1.0$, $\alpha_2 = 2.9$, $\gamma_b=2.8 \times 10^7$. For $\gamma<10^8$, the particles have distributions similar as a broken power law. With $\eta=0.05$, the particles softens at $\gamma\sim10^9$ due to the limit of $\gamma_{\mathrm{max}}$. Differently, the softness of the particles for the other three case above $\gamma\sim10^9$ are due to the energe loss of the synchrotron radiation. 
Fig.~\ref{fig:alpha} shows the effect of $\alpha_1$ and $\alpha_2$ on the SED. The chi-square values associated with the different parameters are also shown in Table~\ref{tab:table1}. In the calculation of the chi-square statistic ($\chi^2$), the number of $\gamma$-ray data points was 24.

\section{summary and discussion}\label{summary and discussion}

PWNe are potential sources of TeV radiation within the Galaxy \citep{2020ChPhC..44f5001A}, where relativistic shock waves are ideal acceleration regions for high-energy particles. We investigated the properties of the multiband nonthermal emission associated with the 100 TeV source LHAASO J1908+0621, and whether it can be originated from the PWN powered by the pulsar PSR J1907+0602 are studied.

We analyzed the $\gamma$-ray emission from the region of LHAASO J1908+0621 using the latest Fermi-LAT data and found that its flux was $(2.02 \pm 0.44) \times 10^{-11} \,\mathrm{erg} \,\mathrm{cm}^{-2} \,\mathrm{s}^{-1}$ with a TS value = 47.58 in the band 30–500 GeV. SrcX is spatially consistent with LHAASO J1908+0621, and the $\gamma$-ray spectrum is very hard with an index of $\sim1.5$ in 30–500 GeV for a single power-law spectrum. A break appears at $\sim 1$ TeV based on the detected fluxes with Fermi-LAT, HAWC and LHAASO.

Based on a one-zone time-dependent model for the multiband nonthermal emission for a PWN, we showed that the detected multiband fluxes associated with LHAASO J1908+0621 can be reproduced from the model with appropriate parameters. The spin-down energy of the pulsar PSR J1907+0602 can  power the particles radiating the detected $\gamma$-rays associated with LHAASO J1908+0621. In model C, the injected particles have an energy spectrum of a broken power-law with indexes of $\alpha_1 = 1.0$ and $\alpha_2 = 2.9$, With a much higher magnetic fraction of $\eta=0.42$ for this nebula compared to the other PWNe in \citet{2014JHEAp...1...31T}, the magnetic field strength is $6.7 \,\mu$G in the emission region of LHAASO J1908+0621. Nevertheless, it is comparable to the inferred value for the wind magnetisation of the PWN Kes 75 \citep{2022arXiv221108816S}.

In this paper, the PWN is assumed to be still in the free expansion phase in which it has not been encountered with the reverse shock. After the encounter, the nebula can be compressed, and the magnetic field in the PWN is significantly strengthened to produce more prominent synchrotron radiation \citep{2009ApJ...703.2051G,2010A&A...515A..20F}.

\section*{Acknowledgements}
This work is supported by NSFC through grants U2031107, 12063004 and 11873042, the grant from Yunnan Province (YNWR-QNBJ-2018-049), the Joint Foundation of Department of Science and Technology of Yunnan Province and Yunnan University  (202201BF070001-020), the Program of Yunnan University (2021Y330) and the National Key R$\&$D Program of China under grant No.2018YFA0404204.
\section*{Data Availability}

The data produced in this paper will be shared on reasonable request to the corresponding author.



\bibliographystyle{mnras}
\bibliography{wky} 




\appendix


\bsp	
\label{lastpage}
\end{document}